# Effect of 0.25 and 2.0 MeV He-ion irradiation on short-range ordering in model (EFDA) Fe-Cr alloys

*Stanisław M. Dubiel[1*], Jan Żukrowski[2] and Yves Serruys[3]*

[1]*AGH University of Science and Technology, Faculty of Physics and Applied Computer Science, al. Adama Mickiewicza 30, 30-059 Kraków, Poland,* [2]*AGH University of Science and Technology, Academic Centre of Materials and Nanotechnology, al. Adama Mickiewicza 30, 30-059 Kraków, Poland,*
[3]*Laboratoire JANNUS, DEN-Service de Recherches de Métallurgie Physique, CEA, Université Paris-Saclay, F-91191, Gif-sur-Yvette, France*

## Abstract

The effects of $He^+$ irradiation on a distribution of Cr atoms in $Fe_{100-x}Cr_x$ (x=5.8, 10.75, 15.15) alloys were studied by $^{57}Fe$ Conversion Electron Mössbauer Spectroscopy (CEMS). The alloys were irradiated with doses up to $12 \times 10^{16}$ ions/cm$^2$ with 0.25 and 2.0 MeV $He^+$ ions. The distribution of Cr atoms within the first two coordination shells around Fe atoms was expressed with short-range order parameters $\alpha_1$ (first-neighbour shell, 1NN), $\alpha_2$ (second-neighbour shell, 2NN) and $\alpha_{12}$ (1NN+2NN). In non-irradiated alloys $\alpha_1>0$ and $\alpha_2<0$ was revealed for all three samples. The value of $\alpha_{12} \approx 0$, i.e. the distribution of Cr atoms averaged over 1NN and 2NN, was random. The effect of the irradiation of the $Fe_{94.2}Cr_{5.8}$ alloy was similar for the two energies of $He^+$ viz. increase of number of Cr atoms in 1NN and decrease in 2NN. Consequently, the degree of ordering increased. For the other two samples, the effect of the irradiation depends on the composition, and is stronger for the less energetic ions where, for $Fe_{89.25}Cr_{10.75}$ alloy, the disordering disappeared and some traces of Cr clustering appeared. In $Fe_{84.85}Cr_{15.15}$ alloy the clustering was clear. In the samples irradiated with 2.0 MeV $He^+$ ions the ordering also survived in the samples with x=10.75 and 15.15, yet its degree became smaller than in the $Fe_{94.2}Cr_{5.8}$ alloy.

* Corresponding author: Stanislaw.Dubiel@fis.agh.edu.pl

## 1. Introduction

Fe-Cr based ferritic steels (FS) such as ODS and ferritic-martensitic (F-MS) steels such as EUROFER constitute an important class of structural materials. Their relevance for numerous industrial and technological applications is a consequence of their desirable swelling, high temperature corrosion and creep resistance properties [1,2]. In these circumstances FS as well as F-MS have been recognized as valuable construction materials for applications in new generations of nuclear power facilities including generation IV fission reactors and fusion reactors as well as for other technologically important plants e.g. high power spallation targets [3–5]. They have, for example, been used for manufacture of such systems as container of the spallation target, fuel cladding or primary vessel. These devices work in-service under extreme conditions such as elevated temperatures and/or long-term irradiation. Under these conditions, the materials experience irradiation damage that can gravely degrade their mechanical properties. On the lattice scale, the radiation produces lattice defects including vacancies, interstitials and dislocations. A redistribution of Fe/Cr atoms subsequently occurs and can produce several microscopic phenomena such as short-range order (SRO), segregation or phase decomposition into Fe-rich ($\alpha$) and Cr-rich ($\alpha'$) phases. All of these effects result in enhanced embrittlement which is highly undesirable. A better understanding of the effects of irradiation on the useful properties of FS/F-MS and underlying mechanisms is an important issue as it may contribute to significantly improve properties of these materials, and, thereby, to extend the operational life of equipment produced therefrom. Fe–Cr alloys, being the basic ingredient of FS/F-MS, have often been used as model alloys for investigations of both physical and technological properties of FS and M-FS steels [6 and references therein]. In laboratory conditions, different projectiles have been used as irradiation media to study irradiation effects in Fe-Cr alloys and/or FS or F-MS. In particular, the effects of neutrons [7,8], protons [9,10], self-ions (Fe, Cr) [11,12], He ions [13,14], Kr ions [15] and electrons [16] have been recently studied. Irradiations were performed under different conditions of temperature, projectile energy, flux and dose. This makes a comparison of the obtained results more difficult as the effects of irradiation depend not only on the type of irradiation but, for a given type, also on the irradiation flux and temperature. To exemplify the former, no $\alpha'$ precipitates were observed in $Fe_{88}Cr_{12}$ alloy irradiated at 300°C with $Fe^+$ ions to 0.6 dpa but they were revealed in the same alloy irradiated with neutrons at the same temperature and with the same dose [17]. Regarding the latter, $\alpha'$ precipitates were found in low dose rate neutron-irradiated samples ($10^{-9} - 10^{-6}$ dpa/s) while they were absent if the dose rate was high ($10^{-4}$ dpa/s) [18]. Similar effects of the dose rate on irradiation hardening were reported for Fe-Cr alloys irradiated with $Fe^+$ ions [19]. These observations illustrate that results obtained for materials irradiated with high dose rates, which is typically the case for *in vitro* studies, may not be simply related to the effects caused by *in vivo* irradiations in which very low dose rates are involved (assuming the same dose of irradiation).

Understanding of irradiation-induced effects can be backed by theoretical calculations such as those recently reported on radiation-induced segregation [20] and radiation-accelerated precipitation [18] in Fe-Cr alloys. In any case further experimental and theoretical studies on these key issues are needed in order to enhance our knowledge and understanding of irradiation-induced phenomena in materials with key engineering significance.

Three model (EFDA/EURATOM) $Fe_{100-x}Cr_x$ alloys (x=5.8, 10.75 and 15.15) were irradiated to different doses with $He^+$ ions of 0.25 and 2.0MeV energy. Samples were investigated using Mössbauer spectroscopy (MS). MS has already proved to be a relevant method for the quantitative investigation of Fe-Cr alloys. Notably issues connected to determination of a distribution of Cr atoms in Fe matrix e. g. [21-27], the solubility limit of chromium [28,29] and the kinetics of $\alpha'$ precipitation [29] can be successfully studied with high precision using MS. The irradiation of Fe-Cr alloys with $He^+$ ions is of interest, as the production of helium occurs during exposure of the various devices produced therefrom to proton and/or neutron irradiation [1]. Its presence deteriorates mechanical properties of steels. In particular, it lowers the critical stress for inter granular structure and also it may bring about a severe decrease of the fracture toughness [30]. Therefore, understanding not only the effects of radiation damage but also the effects of helium on the mechanical properties of FS/F-MS are important topics to be studied in the context of gaining a better understanding of irradiation-induced degradation processes in engineering materials used in various important branches of industry, including nuclear power.

## 2. Experimental

### 2.1. Samples and irradiation

Samples investigated in this study were prepared from model EFDA/ EURATOM master Fe–Cr alloys fabricated in 2007. They were delivered in the form of bars 10.9 mm in diameter, in a re-crystallized state after cold reduction of 70% and then heat-treated for 1h under flow of pure Ar at the following temperatures: 750°C for $Fe_{94.2}Cr_{5.8}$, 800°C for $Fe_{89.25}Cr_{10.75}$ and 850°C for $Fe_{84.85}Cr_{15.15}$, followed by air cooling. For the present study, a slice ~1 mm thick was cut from each bar using a diamond saw, and was subsequently cold-rolled (CR) down to a final thickness of ~30 μm. The samples in form of ~25 mm rectangles were irradiated at the JANNUS multi-ion beam irradiation platform at CEA, Saclay, France with 0.25 and 2.0 MeV $He^+$ ions to a dose of $1.2·10^{17}$ $^4He^+·cm^{-2}$, which is equivalent to radiation damage of 7.5 dpa as calculated by the SRIM code. The irradiation was performed in vacuum at room temperature (290K). The full cascade method was used with the SRIM default values of threshold energies. No ion channeling was expected due to polycrystalline structures of the samples. The irradiation area was circular and had a diameter of 20 mm. Concentration and radiation damage (RD) profiles calculated for Fe-Cr with the SRIM code are displayed in Figs.1 and 2. A pre surface zone of the samples (≲~0.3 μm) accessible for the investigation by the conversion electron Mössbauer spectroscopy (CEMS) is marked by vertical stripes. It can be seen that the sample volume measured with the applied technique was practically free of He, hence the only effect of the irradiation can be of ballistic origin. However, the radiation damage in the investigated zone of the samples significantly depends on the energy of ions viz. it is much stronger in the case of the less energetic projectiles. Consequently, possible differences observed in the samples irradiated with ions of different energies cannot be explained only in terms of a difference in the energy. Hence it is not possible to separate the effect of energy from that of damage level.

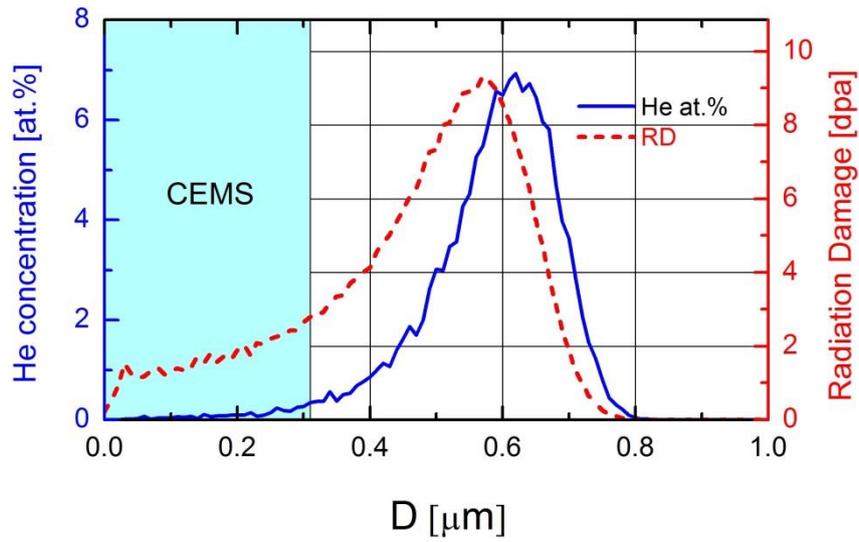

Fig. 1 He-concentration profile and radiation damage calculated by the SRIM code for the Fe$_{89.25}$Cr$_{10.75}$ sample irradiated to the dose of ~7.5 dpa with 0.25 MeV He$^+$ ions vs. depth, *D*. The pre surface zone accessible to the CEMS measurements is marked by a vertical stripe.

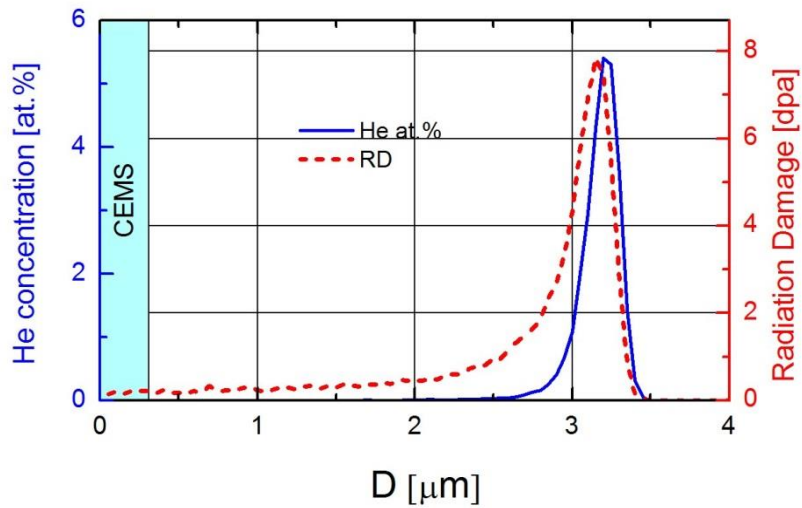

Fig. 2 He-concentration profile as calculated by the SRIM code for the Fe$_{89.25}$Cr$_{10.75}$ sample irradiated to the dose of 7.5 dpa with 2.0 MeV He$^+$. The range of ions is equal to 3.15 μm and the peak concentration ~7.5 at% He. The vertical stripe marks the pre surface zone accessible to the CEMS measurements.

## 2.2. Spectral measurements and analysis

The Mössbauer spectra were measured at room temperature (RT, 290K) by recording conversion electrons (CEMS mode) in a backscattering geometry using a conventional constant acceleration spectrometer and a $^{57}$Co(Rh) source of 14.4 keV gamma-rays with a nominal activity of 3.7 GBq. The measured spectra contain information from a surface/pre surface zone whose thickness is less than ~0.3 μm. The spectra were recorded both on irradiated (IR) as well as on non-irradiated (NIR) surfaces of the samples which were built in a proportional gas flow counter. A He/methane mixture (90:10) was used as the counting gas. The recorded spectra are displayed in Fig. 3. Spectra were analyzed using a two-shell model, assuming the presence of Cr atoms within the two-shells vicinity, 1NN-2NN, of the $^{57}$Fe probe nuclei. Their effect on the hyperfine field (*B*) and on the isomer shift (*IS*) is additive i.e. the following equation holds: *X(n₁,n₂) = X(0,0) + n₁·ΔX₁ +n₂·ΔX₂*, where *X* = *B* or *IS*, *ΔX$_i$* is a change of *B* or *IS* due to one Cr atom situated in 1NN (*i=1*) or in 2NN (*i=2*). The number of Cr atoms in 1NN is indicated by *n₁*, and that in 2NN by *n₂*. Between twelve (the Cr-lowest concentrated sample) and seventeen (the Cr-highest concentrated sample) most significant atomic configurations, *(n₁,n₂)*, were taken into account based on the binomial distribution to fulfill the condition $\sum_{n_1,n_2} P(n_1,n_2) \geq 0.99$. However, the probabilities of the atomic configurations, *P(n₁,n₂)*, were treated as free parameters (their starting values were those calculated from the binomial distribution) in spectral analysis. All spectral parameters such as *X(0,0)*, *ΔX$_i$*, line widths of individual sextets *G1*, *G2* and *G3* and their relative intensities (Clebsch-Gordan coefficients) *C2* and *C3*, were also treated as free parameters (*C1*=1). Very good fits (in terms of a $\chi^2$-test) were obtained with the spectral parameters displayed in Table 1. Their values are in close agreement with the corresponding values reported previously [20-28].

Knowledge of the atomic configurations, *(n₁,n₂)*, and their probabilities, *P(n₁,n₂)*, enabled determination of the average number of Cr atoms in 1NN, $<n_1> = \sum_{n_1,n_2} n_1 P(n_1,n_2)$, in the second, $<n_2> = \sum_{n_1,n_2} n_2 P(n_1,n_2)$ and in both shells, $<n_1+n_2> = \sum_{n_1,n_2} (n_1+n_2) P(n_1,n_2)$. The values of <n₁>, <n₂>, and <n₁+n₂>, were then used for determination of the corresponding SRO parameters, $\alpha_1$, $\alpha_2$, and $\alpha_{12}$, as outlined below.

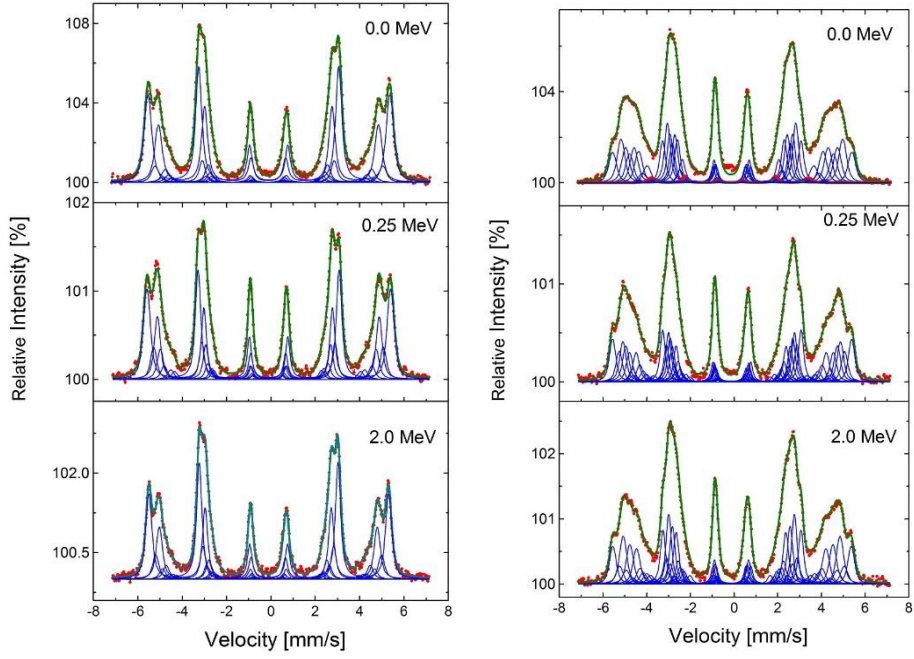

Fig. 3 CEMS spectra recorded at RT on Fe$_{94.2}$Cr$_{5.8}$ (left panel) and Fe$_{84.85}$Cr$_{15.15}$ (right panel) samples irradiated with He$^+$ to the dose of 7.5 dpa. The energy of the ions is indicated by labels. The spectra recorded on non-irradiated surfaces are labelled by 0.0 MeV.

## 3. Results

### 3.1. Short-range order parameters

A distribution of atoms in an alloy (here Fe-Cr) can be quantitatively described using Warren-Cowley short-range order (SRO) parameters, $\alpha_k$. The techniques applied in the present study, i.e. Mössbauer spectroscopy, makes it possible to determine the SRO-parameters for the first, $\alpha_1$, and for the second, $\alpha_2$, nearest-neighbor shells, separately. This enables calculation of the SRO-parameter for both shells, $\alpha_{12}$. In turn, knowledge of the SRO-parameters makes it possible to qualitatively discuss the distribution of Cr atoms in the Fe matrix. The values of $\alpha_k$ ($k$=1,2,12) can be determined using the following equation [23]:

$$\alpha_k = 1 - \frac{<n_k>}{<n_{ok}>} \qquad (1)$$

where $<n_k>$ is the number of Cr atoms in the $k$-th near-neighbour shell around the probe Fe atoms as found from the analysis of the spectra, while $<n_{ok}>$ is the number of Cr atoms in the $k$-th near-neighbor shell calculated assuming their distribution is random i.e. $<n_{01}>=0.08x$, $<n_{02}>=0.06x$, and $<n_{01}+n_{02}>=0.14x$.

### 3.1.1. Effect of He$^+$ energy and chromium concentration

In order to study whether the effects of irradiation depend on the energy of He$^+$ ions and alloy composition, all three samples were irradiated at room temperature (ca. 290K) to a dose of $1.2 \cdot 10^{16}$He$^+$/cm$^2$ (7.5 dpa) with 0.25 and 2.0 MeV ions. The spectra recorded for two samples with the end-member compositions are displayed in Fig. 3. Based on their analysis, as described in section 2.2, and equation (1), values of the SRO-parameters $\alpha_1$, $\alpha_2$ and $\alpha_{12}$ were determined from both the spectra recorded on NIR as well as on IR sides of the samples. The values of the SRO-parameters calculated from the spectra recorded on the non-irradiated surfaces are shown in Fig. 4.

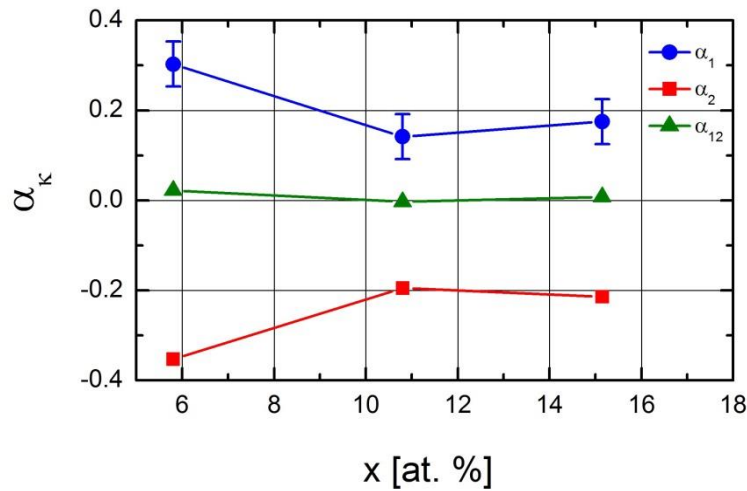

Fig. 4 SRO-parameters $\alpha_1$, $\alpha_2$ and $\alpha_{12}$ versus Cr content, $x$, as determined from the spectra recorded on the non-irradiated surfaces of the Fe$_{100-x}$Cr$_x$ samples. The lines are to guide the eye.

It can be seen that the $\alpha_1$-values are positive for all samples, with the highest value for the lowest-Cr sample. This means, in terms of eq. (1), that the average number of Cr atoms situated in the 1NN-shell around the probe Fe-atoms is smaller than expected for a random distribution. On the other hand, the values of $\alpha_2$ are negative, hence the lattice site occupancy by Cr atoms in the 2NN-shell is overpopulated, as compared with the expectation from a binomial distribution. This means, in terms of the pair interaction potential between Fe and Cr atoms, that the effective potential is repulsive if the separation between these atoms is equal to the 1NN-radius; and it is attractive if the two types of atom are separated from each other by the radius of the 2NN-shell. It must be, however, realized that this effective potential has several contributions including electronic, magnetic, configurational and vibrational ones. It can be also seen that the absolute values of $\alpha_1$ and $\alpha_2$ are significantly larger for the lowest-Cr sample than for the other two samples. This behavior of $\alpha_1$ and $\alpha_2$ indicates an ordering of Cr atoms and, obviously, a degree of the ordering higher for the $Fe_{94.2}Cr_{5.8}$ sample than for the other two samples. The SRO-parameter averaged over the two shells, $\alpha_{12}$, is equal to zero within the error limit. This means that the distribution of Cr atoms as measured within the volume of the 1NN-2NN neighborhood is random.

To figure out the effect of the irradiation on the Cr atom distribution a difference between the corresponding $\alpha_k$-values, $\Delta\alpha_k = \alpha_k(IR) - \alpha_k(NIR) = [\langle n_k(NIR)\rangle - \langle n_k(IR)\rangle]/\langle n_{0k}\rangle$, $k$=1,2,12, determined from the spectra recorded on irradiated (IR) and non-irradiated surfaces, were calculated. They are displayed in Fig. 5 for the irradiation with 0.25 MeV ions and in Fig. 6 for the irradiation with 2.0 MeV ions.

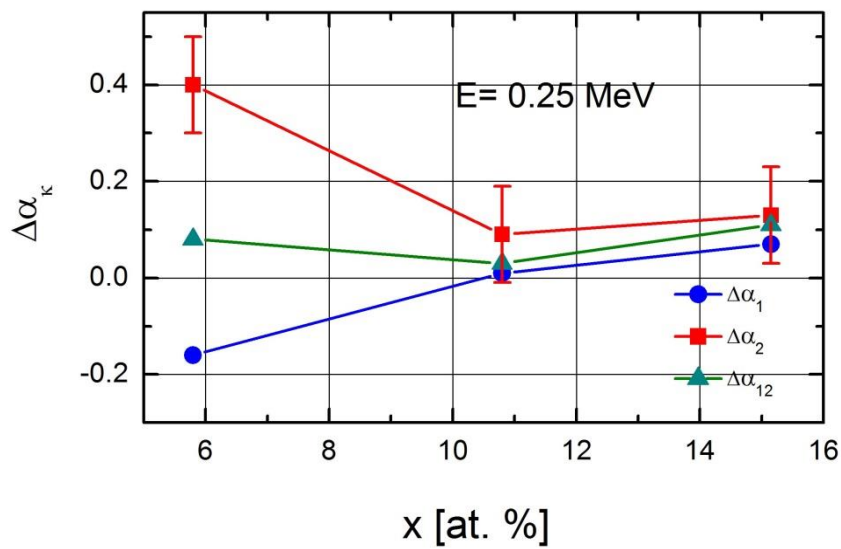

Fig. 5 Dependence of the $\Delta\alpha_k$-parameters on the concentration of chromium, *x*, in $Fe_{100-x}Cr_x$ alloys irradiated with 0.25 MeV $He^+$ ions to the dose of 7.5 dpa. The lines are to guide the eye.

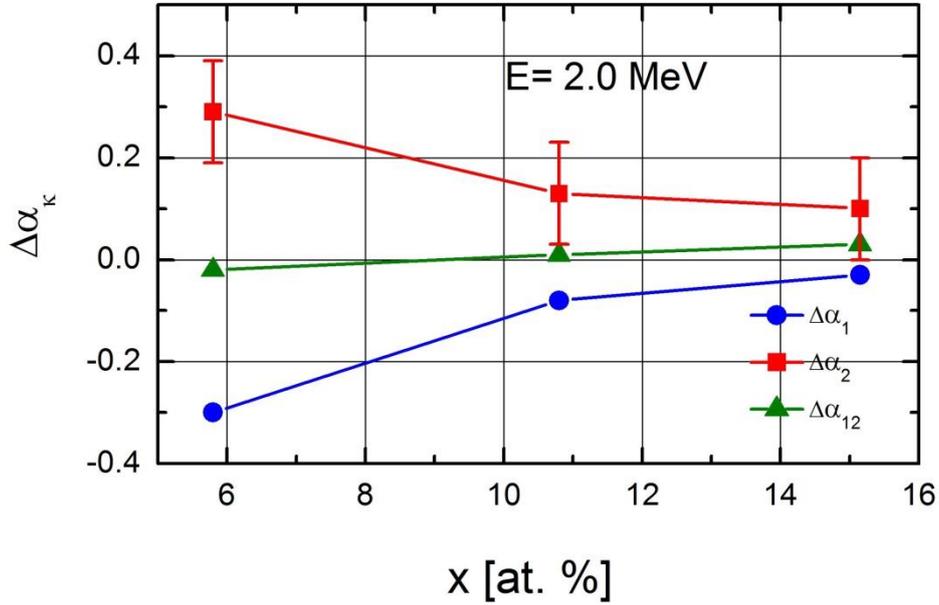

Fig. 6 Dependence of the $\Delta\alpha_k$-parameters on the concentration of chromium, *x*, in $Fe_{100-x}Cr_x$ alloys irradiated to the dose of 7.5 dpa with 2.0 MeV $He^+$ ions. The lines are to guide the eye.

The data displayed in Fig. 5 and 6 give evidence that the effect of the irradiation as seen by the probe $^{57}Fe$ atoms within the ~0.3 μm thick surface/pre surface layer, depends, in general, on the composition of alloys and on the energy of the $He^+$ ions. It is also characteristic of the coordination shell. The strongest effect at both energies is observed for the least Cr-concentrated sample, $Fe_{94.2}Cr_{5.8}$, and the weakest effect for the highest Cr-concentrated alloy i.e. $Fe_{84.85}Cr_{15.15}$. Concerning the former, the irradiation decreased $\alpha_1$ by ~0.16 for 0.25 MeV and by ~0.3 for 2.0 MeV $He^+$ ions, and it increased $\alpha_2$ by ~0.4 for 0.25 MeV and by ~0.3 for 2.0 MeV $He^+$ ions. In other words, under irradiation the number of Cr atoms present in the 1NN-shell increased and that in the 2NN-shell decreased. This effect can be termed as an irradiation-induced ordering of Cr atoms. For the other two samples, the changes in the values of both $\alpha_1$ and $\alpha_2$ are much lower, which means that the redistribution of Cr atoms that had taken place under the irradiation in these samples was minor. Nevertheless, there are some differences to be noticed. In the $Fe_{89.25}Cr_{10.75}$ and $Fe_{84.85}Cr_{15.15}$ samples the ordering ($\alpha_1 < 0$, $\alpha_2 > 0$) survived the irradiation with the 2.0 MeV ions, yet its degree decreased, and in the case of the most Cr-concentrated sample, the $\alpha_{12}$ even became slightly positive i.e. some tiny clustering of Cr atoms may exist. The irradiation with the 0.25 MeV ions destroyed the ordering in the $Fe_{89.25}Cr_{10.75}$ and $Fe_{84.85}Cr_{15.15}$ samples that was present in the un-irradiated samples. The effect is stronger in the $Fe_{84.85}Cr_{15.15}$ sample where both $\Delta\alpha_1$ and $\Delta\alpha_2$ are positive, which means that the number of Cr atoms in both neighbor shells had decreased upon irradiation. Accordingly, the value of $\Delta\alpha_{12}$ is positive, which can be interpreted as an indication of the Cr atom clustering. A similar effect was observed previously for 0.025 MeV $He^+$ projectiles [25,31]. The value of $\Delta\alpha_{12}$ found for the $Fe_{84.85}Cr_{15.15}$ sample but irradiated with 2.0 MeV ions is also positive but its amplitude is ~4 times smaller. This means that the degree of clustering in this sample is correspondingly weaker, i.e. more energetic He ions caused a ballistic disordering of initially ordered Cr atoms.

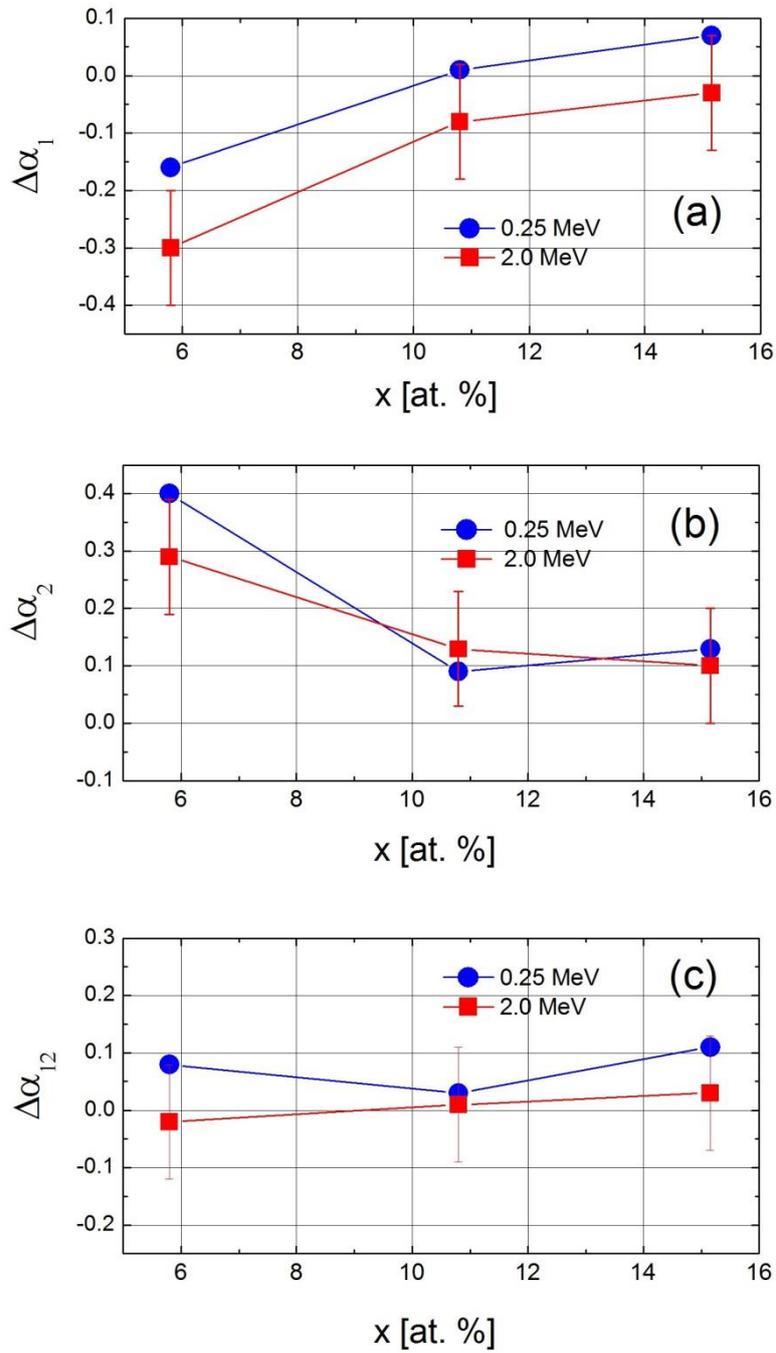

Fig. 7 $\Delta\alpha_k$-parameters for (a) the 1NN-shell, (b) the 2NN-shell, and (c) for the 1NN+2NN-shells versus Cr content, *x*. The lines are to guide the eye.

A comparison of the $\Delta\alpha_k$-values for $k$=1, 2, 12 is presented in Fig. 7. It can be clearly seen that the SRO-parameters are characteristic both of the neighbor shells as well as of the concentration of alloys. For the 1NN-shell they are also characteristic of the energy of the ions. For the 2.0 MeV $\Delta\alpha_1$<0 for all concentrations, but its amplitude strongly decreases with *x.* This means that the number of Cr atoms in the 1NN shell in the irradiated samples is higher than in the non-irradiated samples, but the difference decreases with *x*. The dependence of $\Delta\alpha_1$ for the 0.25 MeV vs. *x* is similar, yet its values are shifted upwards by ~0.1. Consequently, a crossover to a positive value occurs at *x* ≈10 what signifies that upon this irradiation the number of Cr atoms in the 1NN shell was reduced. For the 2NN-shell, the overall trend is rather similar for both energies, i.e. the values of $\Delta\alpha_2$ are positive for all compositions, and the highest for the $Fe_{94.2}Cr_{5.8}$ sample. For the other two samples, it is much lower and practically the same for both energies. This again reflects the weaker effect of the irradiation for the samples with *x* =10 and 15 as compared to the one with *x* =5. The values of $\alpha_{12}$ lie within an error limit close to zero. Possibly a weak clustering of Cr atoms exists for the $Fe_{84.85}Cr_{15.15}$ sample irradiated with less energetic ions. Noteworthy, a similar behavior was previously observed for the alloys of the same origin irradiated to the same dose but with 0.025 MeV $He^+$ ions [32].

The irradiation-induced change of the SRO parameter $\alpha_{12}$ presented in this paper and that reported previously [31] compare pretty well with the recent molecular dynamic simulations (MDS) for disordered $Fe_{100-x}Cr_x$ alloys [33], as far as the alloys with *x*=10 and 15 are concerned viz. $\alpha_{12}$ is unaffected by the irradiation for the former and becomes positive for the latter *x*-value. For the least Cr-concentrated sample i.e. *x*=5, the measured $\alpha_{12}$ remained unchanged for the alloy irradiated with the 2.0 MeV ions, but increased in the case of the 0.25 MeV ions. For this alloy the MDS predicted negative value for $\alpha_{12}$. The comparison should be, however, taken with caution, because it was assumed in the MDS that the distribution of Cr atoms was random in the non-irradiated samples i.e. $\alpha_{12}$ =0. In our case, $\alpha_{12}$ was also equal to zero – see Fig. 4, but the distributions of Cr atoms over the 1NN and 2NN shells were not random ($\alpha_1$ > 0, $\alpha_2$ < 0). It should be also noticed that the MDS were performed for the overlapping of single 5 keV displacement cascade events, therefore its relevance to our study is not straightforward. In any case, the results reported in this paper are, to our best knowledge, the first that give evidence on clustering of Cr atoms in a $Fe_{84.85}Cr_{15.15}$ alloy irradiated with $He^+$ ions. This observation may contribute to a better understanding of mechanisms underlying irradiation-induced effects by He ions in Fe-Cr alloys and produced therefrom nuclear structural materials. In the available literature there are numerous reports on the issue e. g. [13, 14, 34-41]. The reported results have not permitted to obtain the full understanding of the underlying mechanism and the observed effects like enhanced hardening, embrittlement and swelling are still a matter of discussion and even controversy. This situation has many reasons, and first of all a lack of systematic studies in terms of various parameters that can affect mechanical and structural properties of these materials. It is well known that such conditions of irradiation like: temperature, dose, fluence, radiation damage, energy of He ions and composition of materials have to be taken into account. In particular, it was revealed in previous studies that the lower the temperature of the irradiation the higher the degree of hardness [36, 38]. The dose and the radiation damage must be high enough to result in the increased hardening [39, 41]. The effect of the irradiation was also revealed to depend on the irradiated material [39, 40]. The observed mechanical effects of the He-irradiation are most frequently explained in terms of formation of He-bubbles, especially small ones, and also changes in the microstructure induced by the displacement damage [35, 36]. However, Ullmaier and Camus concluded their study that the increase in yield stress following the He-irradiation was solely due to displacement induced defects but not to the presence of He itself [34]. It should be noticed here that the above-discussed changes in the microstructure of the He-irradiated materials were mostly investigated using the transmission electron microscopy (TEM) and, rarely, by small angle neutron scattering (SANS) e.g. [36]. These methods do not allow detecting changes in a distribution of Cr atoms that, as we have shown in this study, occur upon He-ion irradiation. In other words, in order to properly understand mechanisms responsible for the irradiation-caused changes in the mechanical properties of nuclear materials one has also to take into account changes in the distribution of atoms in the matrix. It is well known that clustering of Cr atoms in Fe-Cr alloys and in Fe-Cr based steels results in an enhanced embrittlement, hence the redistribution cannot be neglected.

## 3.2. Changes in local concentration of Cr

The observed irradiation-induced changes in the distribution of Cr atoms within the 1NN-2NN volume around the probe Fe atoms can be also expressed in terms of underlying changes in the local concentration of Cr, $x_k$, defined as follows [26]:

$$x_k(at.\%) = \frac{<n_k>}{M} 100 \qquad (2)$$

Where $M$=8, 6, 14 for $k$=1, 2, 12, respectively, stays for the maximum number of atoms in 1NN, 2NN and 1NN+2NN shells.

The $x_k$ – values obtained based on eq. (1) are displayed in Table 1. It can be readily noticed that, as a rule, the $x_1$-values are systematically smaller and those of $x_2$ systematically greater than the corresponding average values. To extract the effect of irradiation a difference in $x_k$, $\Delta x_k = x_k(IR) - x_k(NIR)$, has been calculated for all $k$=1, 2, 12 and both energies. The output of these calculations is presented in Figs. 7 and 8 for 0.25 MeV and 2.0 MeV, respectively.

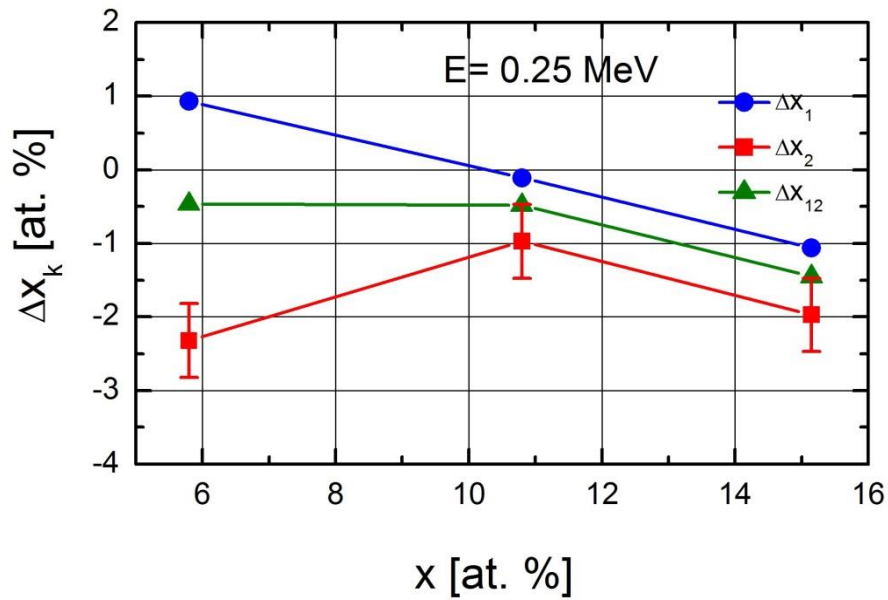

Fig. 8 Dependence of $\Delta x_k$ ($k$=1, 2, 12) on Cr concentration, $x$, for the samples irradiated with 0.25 MeV He$^+$ ions. The lines are to guide the eye.

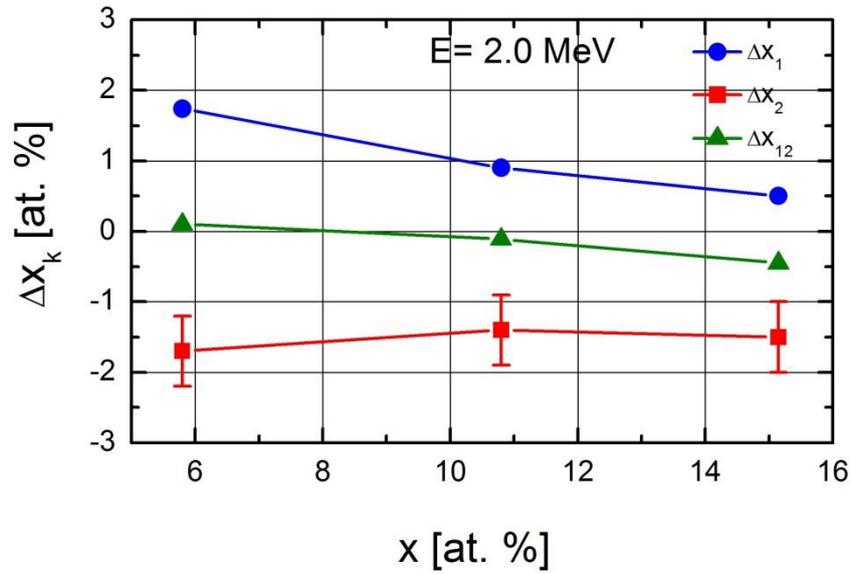

Fig. 9 Dependence of $\Delta x_k$ ($k$=1, 2, 12) on Cr concentration for the samples irradiated with 2.0 MeV He$^+$ ions. The lines are to guide the eye.

Here, the effect of 0.25 MeV projectiles and those of 2.0 MeV is seen in terms of the underlying changes in the local concentration of Cr in particular shells, as well as in both shells. Concerning the less energetic ions, the effect strongly depends on the sample. In the least Cr-concentrated one, the content of Cr in the 1NN shell increased by ~ 1 at. % and that in the 2NN shell decreased by ~2 at. %. This means that the degree of the ordering was enhanced by the irradiation. In the Fe$_{89.25}$Cr$_{10.75}$ sample, the local concentration of Cr in the 1NN shell remained unaffected while in the 2NN shell it is decreased by ~1 at. %. Finally, in the most Cr-concentrated sample, the irradiation resulted in a decrease of the local Cr content in both shells viz. by ~1 at. % in the 1NN and by ~2 at. % in the 2NN. On average, the irradiation caused ~0.5 at. % decrease of Cr within the 1NN-2NN shells volume around the probe Fe atoms for x=5.8 and 10.75 at. % Cr and ~1.5 at. % decrease in the Fe$_{84.85}$Cr$_{15.15}$ sample. This effect can be termed as the irradiation-induced clustering of Cr atoms. The corresponding behavior revealed in the studied samples irradiated with 2.0 MeV He ions turned out to be different. Namely, for the 1NN shell there is an increase of the local Cr content for all samples and it ranges between ~2 at. % for the Fe$_{94.2}$Cr$_{5.8}$ sample and ~0.5 at. % for the Fe$_{84.85}$Cr$_{15.15}$ sample. For the 2NN shell there is ~1.5 at. % decrease found for all three samples. On average, i.e.

within the 1NN-2NN volume, no change was found for x=5.8 and 10.75 while a weak decrease exists for x=15.15. In other words, the ordering of Cr atoms observed in non-irradiated samples was enhanced by the irradiation with the 2.0 MeV He ions. The degree of the enhancement, however, decreases weakly with x, and, for the most Cr-concentrated sample a weak clustering of Cr atoms possibly occurs.

The results displayed in Figs. 5 through 9 clearly show that the $Fe_{94.2}Cr_{5.8}$ sample behaves differently than the other two ones. Namely, the applied irradiation caused in this sample the greatest changes in the distribution of Cr atoms. Interestingly, the greatest increase in the hardness caused by an $Fe^+$ ion irradiation was also observed in a $Fe_{94.2}Cr_{5.8}$ sample (having the same origin as ours) [19]. Theoretical calculations showed that the mixing energy in random $Fe_{100-x}Cr_x$ alloys is negative for $x \leq \sim10$ and has its minimum at ~5at% Cr [32], which correlates with our findings.

### 3.3. Change in the magnetic texture

The knowledge of a relative intensity of the 2$^{nd}$/5$^{th}$ line, *C2/C3,* can be used to determine an average angle between the direction of the γ-rays (in this case perpendicular to samples surface) and that of the magnetization vector, Θ. For this purpose the following equation can be used:

$$C2/C3 = \frac{4\sin^2\Theta}{1+\cos^2\Theta} \qquad (3)$$

The Θ-values obtained from equ. (3) and the C2/C3 data displayed in Table 1 are shown in Fig. 10.

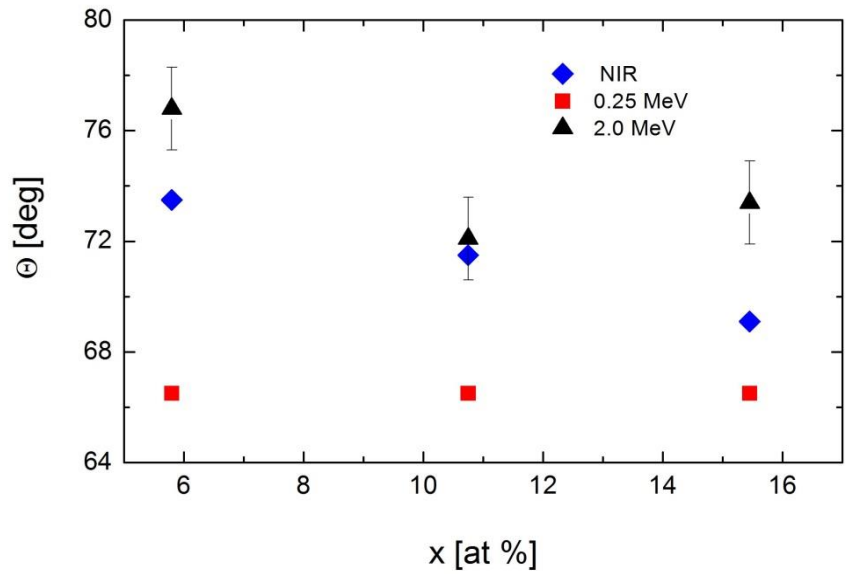

Fig. 10 Average angle, $\Theta$, between the $\gamma$-rays (normal to the samples surface) and the magnetization vector within the presurface zone of the investigated alloys.

It is clear that the $\Theta$-values are characteristic of the sample. In the non-irradiated samples $\Theta$ linearly decreases with *x.* However, this behavior does not necessarily reflect the effect of composition, but it may be due to a various degree of deformation caused by cold rolling of the samples [26]. In order to extract the effect of the irradiation we calculated the difference in $\Theta$, $\Delta\Theta=\Theta(IR)- \Theta(NIR)$, and the result is displayed in Fig. 11.

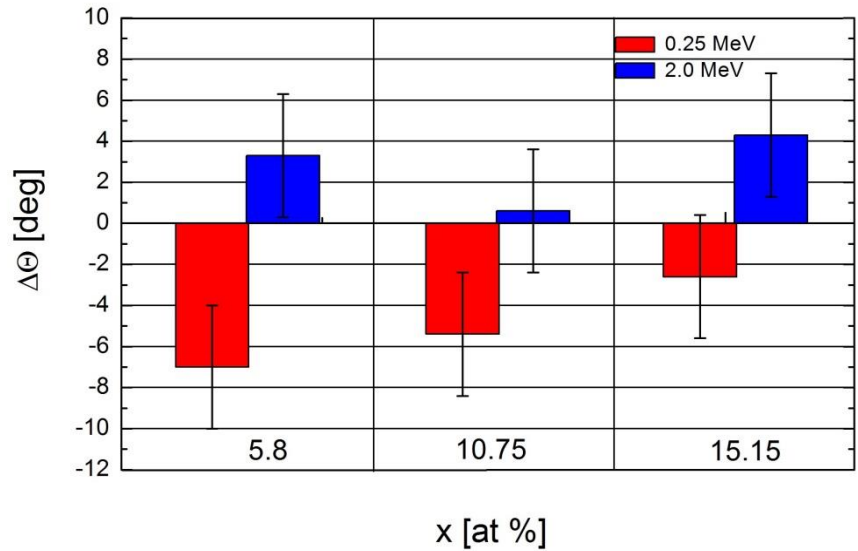

Fig. 11 The difference in Θ, ΔΘ=Θ(IR)- Θ(NIR), vs. Cr concentration, *x*, for the studied samples irradiated with He$^+$ ions of different energy. Error bars are indicated.

The irradiation-induced change of Θ evidently depends on the energy of the ions. Whereas the less energetic ions decrease Θ in a way that seems to depend on the composition, the more energetic ions act in the opposite way, i.e. they increase Θ with the minimum at *x* = 10.75. Although the revealed changes in Θ are rather small, evidently the effect of the 0.25 MeV ions is opposite to the one due to the 2.0 MeV ions.

## 4. Conclusions

The following conclusions can be drawn based on the results obtained in this study:

(1) The population of Cr atoms in the first-neighbour shell (1NN) around the probe Fe atoms in the non-irradiated alloys is lower ($\alpha_1$>0) and in the second-neighbour shell (2NN) is higher ($\alpha_2$<0) than expected from the binomial distribution.

(2) Upon He$^+$ ion irradiation, the degree of the ordering increased, i.e. $\alpha_1$ became more negative and $\alpha_2$ more positive in the least Cr-concentrated sample whereas it remained quasi unchanged in the other two alloys.

(3) Irradiation with the 0.25 MeV ions proved to be more effective as far as the redistribution of Cr atoms is concerned, and, in particular, it caused a clustering of Cr atoms in the most Cr-concentrated sample.

(4) Clustering of Cr atoms was not found in the samples irradiated with the 2.0 MeV ions, but the ordering observed in the non-irradiated samples was enhanced. The degree of the enhancement was decreasing with the Cr content.

(5) Irradiation with the He ions changed the magnetic texture within the investigated pre surface zone viz. the magnetization vector rotated towards the normal to the surface by ~3-6°, depending on the composition, in the 0.25 MeV irradiated samples, while in the 2.0 MeV irradiated samples it rotated towards the samples' surface by up to ~4°.

**Acknowledgements**

This work has been carried out within the framework of the EUROfusion Consortium and has received funding from the Euratom research and training programme 2014-2018 under grant agreement No 633053. The views and opinions expressed herein do not necessarily reflect those of the European Commission. It was also supported by The Ministry of Science and Higher Education, Warszawa, Poland. This work was supported the Ministry of Science and Higher Education (MNiSW), Warsaw, Poland. The irradiation was done at JANNUS-Saclay (Joint Accelerators for Nanoscience and NUclear Simulation), CEA, France and supported by the French Network EMIR.

**Figures captions:**

Fig. 1 He-concentration profile and radiation damage calculated by the SRIM code for the Fe$_{89.25}$Cr$_{10.75}$ sample irradiated to the dose of ~7.5 dpa with 0.25 MeV He$^+$ ions vs. depth, *D*. The pre surface zone accessible to the CEMS measurements is marked by a vertical stripe.

Fig. 2 He-concentration profile as calculated by the SRIM code for the Fe$_{89.25}$Cr$_{10.75}$ sample irradiated to the dose of 7.5 dpa with 2.0 MeV He$^+$. The range of ions is equal to 3.15 μm and the peak concentration ~7.5 at% He. The vertical stripe marks the pre surface zone accessible to the CEMS measurements.

Fig. 3 CEMS spectra recorded at RT on Fe$_{94.2}$Cr$_{5.8}$ (left panel) and Fe$_{84.85}$Cr$_{15.15}$ (right panel) samples irradiated with He$^+$ to the dose of 7.5 dpa. The energy of the ions is indicated by labels. The spectra recorded on non-irradiated surfaces are labelled by 0.0 MeV.

Fig. 4 SRO-parameters $\alpha_1$, $\alpha_2$ and $\alpha_{12}$ versus Cr content, *x*, as determined from the spectra recorded on the non-irradiated surfaces of the Fe$_{100-x}$Cr$_x$ samples.

Fig. 5 Dependence of the $\Delta\alpha_k$-parameters on the concentration of chromium, *x*, in Fe$_{100-x}$Cr$_x$ alloys irradiated with 0.25 MeV He$^+$ ions to the dose of 7.5 dpa.

Fig. 6 Dependence of the $\Delta\alpha_k$-parameters on the concentration of chromium, *x*, in Fe$_{100-x}$Cr$_x$ alloys irradiated to the dose of 7.5 dpa with 2.0 MeV He$^+$ ions.

Fig. 7 $\Delta\alpha_k$-parameters for (a) the 1NN-shell, (b) the 2NN-shell, and (c) for the 1NN+2NN-shells versus Cr content, *x*. The lines visualize trends.

Fig. 8 Dependence of $\Delta x_k$ ($k$=1, 2, 12) on Cr concentration, $x$, for the samples irradiated with 0.25 MeV He$^+$ ions. The lines are to guide the eye.

Fig. 9 Dependence of $\Delta x_k$ ($k$=1, 2, 12) on Cr concentration for the samples irradiated with 2.0 MeV He$^+$ ions. The lines are to guide the eye.

Fig. 10 Average angle, $\Theta$, between the $\gamma$-rays (normal to the samples surface) and the magnetization vector within the presurface zone of the investigated alloys.

Fig. 11 The difference in $\Theta$, $\Delta\Theta=\Theta(IR)- \Theta(NIR)$, vs. Cr concentration, $x,$ for the studied samples irradiated with He$^+$ ions of different energy. Error bars are indicated.